# Comparison of magnetic vector and total scalar potential formulations for finite-element modeling of dipole magnet with COMSOL Multiphysics®.

ALEXANDER CHERVYAKOV

Laboratory of Information Technologies, Joint Institute for Nuclear Research,
Joliot Curie 6, 141980 Dubna, Russia
*acher@jinr.ru*

Abstract

Finite-element simulations of magnetostatic fields are performed in terms of magnetic vector and total scalar potentials and compared for purpose of modeling the accelerator magnets. The potentials represent the unknown variables associated with the A– and V– formulations of magnetostatics to describe the magnetic fields. The simulations are carried out with a single software package, the COMSOL Multiphysics®, where both formulations are implemented. The numerical performance of these methods is illustrated with the model example of a superconducting dipole magnet recently developed for operation in the isochronous cyclotron SC200. The results of calculations are analyzed and compared in terms of the relevant FEM parameters accounting for performance of computation as well as the computational cost. We show in particular that the use of scalar potential as compared to its vector counterpart substantially reduces the number of degrees of freedom, the usage of computer memory and the computational time for a similar relative error.

Introduction

Finite-element modeling of accelerator magnets is usually confronted by complexities of geometries, nonlinearities of materials and high-quality requirements for magnetic fields. For such a modeling, attaining a sufficient accuracy at affordable computational cost is a tedious task. Several software packages based on different formulations of the magnetostatics are presently available to optimize the performance of calculations [1]. The magnetostatic simulations with COMSOL Multiphysics® [2], for example, deploy mainly the magnetic vector potential, which is set as unknown variable describing the magnetic field (A–formulation). However, despite the great efficiency of this formulation for simulations of the accelerator magnets, the use of vector potential for the whole problem domain including the extended current-free regions substantially increases the total number of degrees of freedom (DOFs) and the associated computational cost. Alternatively, the total scalar potential (V–formulation) is now made also available for computations with this software after introducing the cut surfaces assigned with the potential jumps for the uniqueness of its definition [2]. It becomes therefore possible to compare the numerical efficiency of the both methods within the same software package. For this comparison, we utilize the model of a superconducting dipole magnet [3] operated in the isochronous cyclotron SC200, which is recently developed for medical applications [4,5]. To benchmark this model with the two methods, each coil of the magnet is approximated by the line currents located at determined conductor positions inside the coil for computations with magnetic vector potential. The associated potential surfaces bounded by these loops are then constructed and implemented for computations with total scalar potential. This provides the uniform geometry of the magnet for both computational methods. The benchmark model is solved as a finite-element problem. The quantity of interest is the magnetic flux in the aperture region in between the poles of the magnet. To calculate this quantity, the two studies relied on the same model geometry are performed. The first focuses on the performance of the magnetic vector potential formulation. The second repeats a similar calculation using the total scalar potential formulation. In both formulations, the magnetic fields are determined as spatial derivatives of the potentials. This may result in the local inaccuracies that have to be corrected by a postprocessing techniques. The numerical performance of the both potential formulations is analyzed and compared in terms of the relevant FEM parameters such as mesh quality, number of degrees of freedom (DOFs), polynomial order of the discretization and accuracy of solutions. The consumption of memory and time is also taken as a decisive factor for this comparison. In particular, we show that the use of scalar potential as compared to its vector counterpart substantially reduces the number of degrees of freedom, the usage of computer memory, and especially, the computational time, while a similar accuracy of magnetic fields is attained with the both methods. In more details, simulations of the magnetic system for the isochronous cyclotron SC200 with COMSOL Multiphysics® is described in review article [6].

1. Benchmark model of a superconducting dipole magnet

For these calculations, we implement the model of a superconducting dipole magnet [3] specifically designed to produce the focusing magnetic fields for operation of the isochronous cyclotron, whose compact model SC200 [4,5] is aimed at creating proton beams with the energies of 200 MeV for medical applications. The geometry of the magnet is shown in Fig.1.

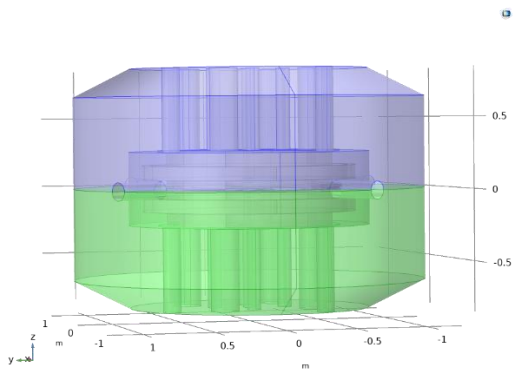

Fig.1: Dipole magnet of the isochronous cyclotron SC200 for proton therapy.

The magnet is capable of producing magnetic fields in accordance with conditions that are necessary for sustainable cyclotron operation [7]. A special care in particular is taken to ensure the field quality in the aperture region where the beam is injected and extracted. In this region, the magnetic field should be as homogeneous as possible for precise beam focusing. The field should be also cyclic along the azimuth to provide the vertical focusing of the particle orbit and increasing in the radial direction to compensate the relativistic mass change and maintain the isochronism between the orbital frequency of trajectory and the radio frequency of accelerating electric field.

The either pole of the magnet consists of the yoke, four spiral sectors, a superconducting coil and also includes smaller constituents such as injector and beam headers [3]. All these parts except for the coil are made out of the ferromagnetic materials with nonlinear magnetization. The magnet is assumed to be magnetically insulated by a surrounded air.

Each coil is driven by the DC current, whose value amounts effectively to $725500$ $[A \times turn]$. The coil geometry represents the axisymmetric hollow cylinder formed by rotation around the $z$ axis of a rectangle lying in the plane $(z, x)$ with center located at the point (0.1145 m, 0.7311 m). The rectangle has the length of 0.096 m and the width of 0.0582 m. The values of internal and external coil radius amount therefore to 0.7020 m and 0.7602 m, respectively.

The model geometry consists of the magnet inside a sphere which is surrounded by the boundary layer, whose thickness can be scaled towards infinity to represent virtually the infinite element domain. The actual size of the model is reduced by a factor eight by exploiting the symmetries of the magnet in order to decrease the computational cost. The mirror symmetry with respect to median plane is used to deploy only a single pole of the magnet surrounded by a semi-sphere with the second pole being accounted for by means of symmetry conditions. The remaining geometry of a single pole is then reduced further by using the four-fold axial symmetry with respect to rotations around third axis. The symmetry is realized by applying the magnetic insulation boundary conditions on symmetry planes. This allows us to implement only the 1/4 part of the pole geometry to simulate the magnetic fields of the whole magnet. The resulting geometry of the magnet is shown in Fig.2.

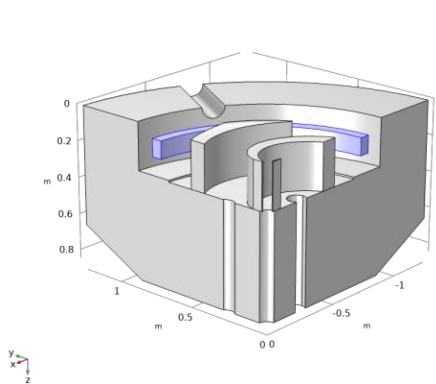

Fig.2: The 1/8 part of the magnet geometry for computations of the reference magnetic field.

The geometry in Fig.2 is implemented in COMSOL Multiphysics® to obtain the reference magnetic field using the Magnetic Fields interface of the AC/DC module [2]. In this interface, the field is computed in terms of the magnetic vector potential. The coil is modeled by means of the add-on Coil future as a circular multi-turn coil driven by DC current, whose current-carrying density is averaged over the coil cross-section.

For direct comparison of computations in terms of the magnetic vector as well as the total scalar potentials, the coil is however modeled within its present geometry differently. For this modeling, the infinitely thin current-carrying loops are implemented for computations with magnetic vector potential and the associated potential surfaces bounded by these loops for computations with total scalar potential. This provides the uniform geometry of the magnet for benchmarking with both computational methods.

Specifically, each coil is represented as a set of a certain number $N$ of the infinitely thin current-carrying loops arranged in parallel at equally small distances apart from each other. Each loop carries the equal fraction $1/N$ of the net current. The size, position and number of loops have been preliminary optimized and validated against the reference magnetic field [8]. The resulting geometry of the model for field computations in terms of the magnetic vector potential is shown in Fig.3.

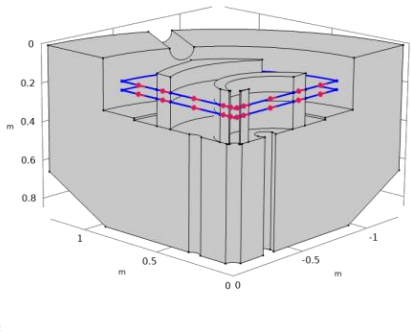

Fig.3: The 1/8 part of the benchmark geometry for field computations with magnetic vector potential.

The geometry in Fig.3 is implemented in COMSOL Multiphysics® using the Magnetic Fields interface of the AC/DC module [2]. The coil is modeled as a bunch of the infinitely thin current-carrying loops using the add-on Edge Current future. The magnetic field is computed in terms of the magnetic vector potential and validated against the reference magnetic field.

The geometry of the model with current loops is then adapted for field computations with total scalar potential. To this end, all loops are spanned by the arbitrary surfaces to cut the nonconducting regions enclosed by currents. Each cut surface whenever traversed in the direction of its normal is assigned with a potential difference, whose value is equal to the fraction $1/N$ of the coil total current. The potential jumps account formally for the currents flowing in the loops. The resulting geometry of the model for field computations in terms of the total scalar potential is shown in Fig.4.

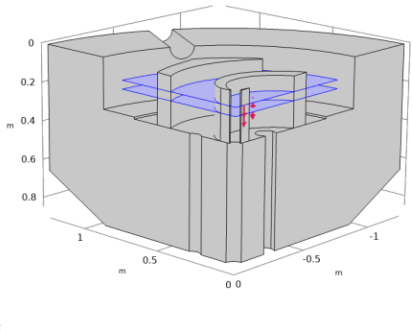

Fig.4: The 1/8 part of the benchmark geometry for field computations with total scalar potential.

The geometry in Fig.4 is implemented in COMSOL Multiphysics® using the Magnetic Fields, No Currents interface of the AC/DC module [2] to compute the field in terms of the total scalar potential. The potential surfaces associated with current loops are modeled by means of the add-on Magnetic Scalar Potential Discontinuity feature.

For coil total current indicated as $I = 725500\ [A \cdot turn]$, the current flowing in each loop and the potential jump over each cut surface are set to $I/N$, where $N$ is the number of either turns, or cuts. The actual value of $N$ is varied starting with $N = 2$ in these calculations.

2. Modeling frameworks

*2.1 Equations of magnetostatics*

For static magnetic fields caused by continuous currents, the basic equations are defined by laws due to Ampere and Gauss, respectively as [9]

$$\nabla \times \boldsymbol{H} = \boldsymbol{j}, \tag{1}$$
$$\nabla \cdot \boldsymbol{B} = 0, \tag{2}$$

where the vectors $\boldsymbol{B}$, $\boldsymbol{H}$ and $\boldsymbol{j}$ represent the magnetic flux density, magnetic field strength and current density, respectively. For neighboring regions of different materials, the continuity of the normal component of $\boldsymbol{B}$ and tangential component of $\boldsymbol{H}$ at boundary interface are ensured by the following conditions

$$\boldsymbol{n} \cdot (\boldsymbol{B}^1 - \boldsymbol{B}^2) = 0, \tag{3}$$
$$\boldsymbol{n} \times (\boldsymbol{H}^1 - \boldsymbol{H}^2) = \boldsymbol{0}, \tag{4}$$

where superscripts refer to material mediums 1 and 2, respectively. In addition to equations (1) – (4), the magnetic flux density $\boldsymbol{B}$ and the field strength $\boldsymbol{H}$ are subjected to the following constitutive relation accounting for the magnetic properties of materials

$$\boldsymbol{B} = \mu \cdot \boldsymbol{H} = \mu_0 \cdot (\boldsymbol{H} + \boldsymbol{M}), \tag{5}$$

with $\boldsymbol{M}$ being the magnetization and $\mu_0$ the permeability of a free space. The permeability $\mu(H) = \mu_0 \mu_r(H)$ as well as the relative permeability $\mu_r(H)$ are of the material specific, and therefore, of the field dependent for materials with nonlinear magnetization. According to equation (5), there are three unknowns for system of four scalar equations (1) and (2) to determine the magnetostatic fields.

*2.2 Magnetic vector potential $\boldsymbol{A}(x, y, z)$ (A – formulation)*

The standard approach allowing for unique solution to system of the magnetostatic equations consists of exploiting the magnetic vector potential $\boldsymbol{A}(x, y, z)$ together with the Coulomb gauge $\nabla \cdot \boldsymbol{A} = 0$ to define the magnetic flux density as $\boldsymbol{B} = \nabla \times \boldsymbol{A}$. This solves identically the scalar equation (2) for the magnetic flux conservation law, while the vector equation (1) for the Ampere's law becomes

$$\nabla \times \left(^1/_\mu \nabla \times \boldsymbol{A}\right) = \boldsymbol{j}. \tag{6}$$

To take into account the nonlinear magnetic properties of materials, the constitutive relation (5) is rewritten first in the form $\boldsymbol{H} = \left(^1/_\mu\right)\boldsymbol{B}$, and then more explicitly, as

$$\boldsymbol{H} = f(|\boldsymbol{B}|) \cdot \frac{\boldsymbol{B}}{|\boldsymbol{B}|}. \tag{7}$$

In this formulation, the mirror symmetry of the dipole magnet is ensured by enforcing the continuity of both the normal component of the current density and the tangential component of the magnetic flux density across the midplane as

$$\boldsymbol{n} \times \boldsymbol{H} = 0. \tag{8}$$

In addition, the four-fold axial symmetry of the magnet is accounted for by imposing the continuity of the tangential component of the current density and the normal component of the magnetic flux density at each symmetry plane

$$\boldsymbol{n} \times \boldsymbol{A} = 0. \tag{9}$$

The A – formulation is implemented in the Magnetic Fields interface of the AC/DC module, where the boundary conditions (8) and (9) are accounted for by in-build nodes of the Perfect Magnetic Conductor and Magnetic Insulation, respectively.

### 2.3 Total scalar potential (V – formulation)

Alternatively, the total scalar potential $V_m(x, y, z)$ can be used to solve equation (1) for Ampere's law in the current-free regions ($\boldsymbol{j} = \boldsymbol{0}$) as $\boldsymbol{H} = -\nabla V_m$. The consistency of solution is ensured by introducing the cuts $\alpha = 1, \ldots, N$, whose surfaces are restricted by the current-carrying loops $I_\alpha$. Furthermore, each cut surface is assigned with a potential jump equal to the current flowing in the surrounding loop

$$\Delta V_{m\alpha} = V_{m\alpha}^+ - V_{m\alpha}^- = I_\alpha, \quad \alpha = 1, \ldots, N. \tag{10}$$

By this means, the total scalar potential becomes a single-valued function allowing to define the magnetic fields everywhere in the simply-connected regions enclosed by currents. In terms of scalar potential, the Ampere's law is then consistently satisfied, whereas the equation (2) for magnetic flux conservation reads

$$-\nabla \cdot (\mu \cdot \nabla V_m) = 0, \tag{11}$$

A nonlinear form of the constitutive relation (5) is now written as

$$\boldsymbol{B} = f(|\boldsymbol{H}|) \cdot \frac{\boldsymbol{H}}{|\boldsymbol{H}|}. \tag{12}$$

In this formulation, the mirror symmetry of the dipole magnet is accounted for by imposing the zero potential boundary condition at median plane

$$V_m = 0. \tag{13}$$

In addition, the four-fold axial symmetry of the magnet is ensured by enforcing the continuity of the normal component of the magnetic flux density at each symmetry plane

$$\boldsymbol{n} \cdot \boldsymbol{B} = 0. \tag{14}$$

The V – formulation is implemented in the Magnetic Fields, No Currents interface of the AC/DC module, where the boundary conditions (13) and (14) are accounted for by in-build nodes of the Zero Magnetic Scalar Potential and Magnetic Insulation, respectively.

For FEM application, the basic equations are represented in the weak form [1,2]. The magnet geometry is approximated by a finite-element mesh. In the presence of different materials, each mesh element can only be assigned with a single material type. The interfaces between neighboring elements of different materials are attributed to the above indicated boundary conditions. Both potentials are then discretized on the finite-element mesh and approximated by using either nodal, or edge shape functions. The nodal shape functions are used for approximation of the total scalar potential. The potential discontinuities across the triangulated cut surfaces are accounted for by constructing each triangle as the common face of a couple of neighboring tetrahedra located at opposite sides of the cut surface. The duplicated nodes are then assigned with the value of the potential difference which is equal to corresponding fraction of the net current flowing in the surrounding loop. For approximation of the magnetic vector potential, on the other hand, the edge shape functions are used to automatically maintain the continuity of its tangential component at interfaces between the mesh elements. The divergence-free Coulomb condition is identically fulfilled by construction of the curl elements. The magnetic fields are determined element-wise as first derivatives of discretized potentials. Accordingly, their order of the discretization becomes one less by using the polynomial shape functions.

3. Results of the field modeling

The quantity of interest is the magnetic flux in the aperture region between the poles of the magnet. The purpose of simulations is to determine the azimuthal as well as the radial magnetic flux density distributions along the midplane of the magnet. The aperture fields are calculated with the precision capable of detecting the harmonic distortion factors of the order of $10^{-4}$ of the field magnitude.

*3.1 Calculation of azimuthal field distributions*

Figures 5 – 7 present the field maps on median plane and its azimuthal profiles calculated for dipole magnet by the finite-element method using COMSOL Multiphysics®. The reference distributions are shown in Fig.5 and the distributions calculated with magnetic vector and total scalar potentials in Fig.6 and Fig.7, respectively. Figure 8 shows a comparison of the azimuthal field profiles calculated by the two methods with each other and against the reference magnetic field.

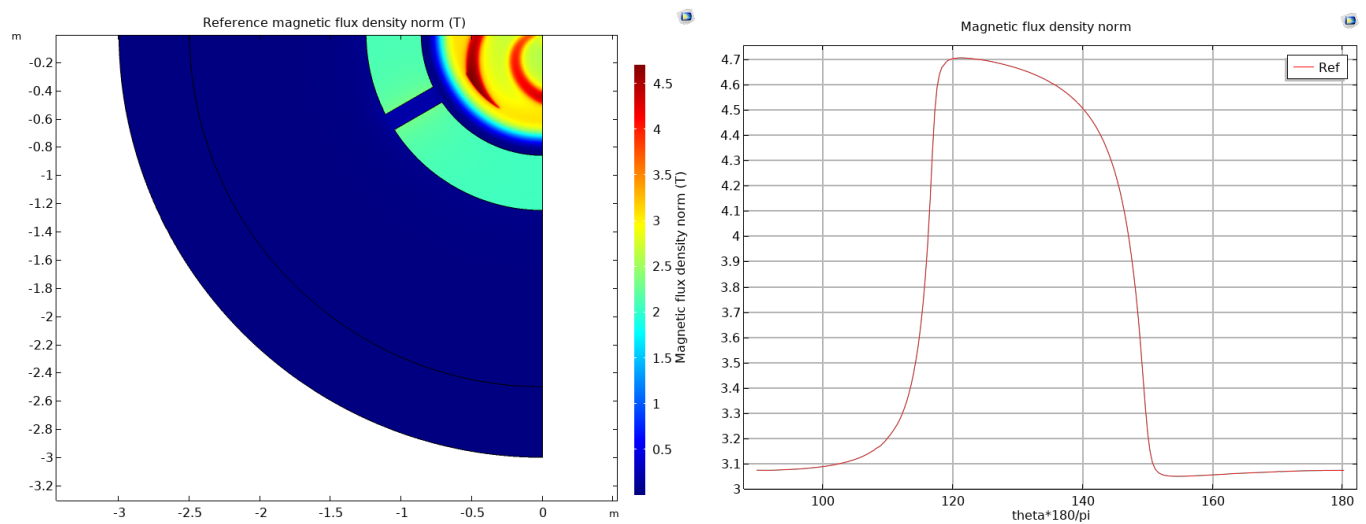

Fig.5: Reference field distributions for dipole magnet.

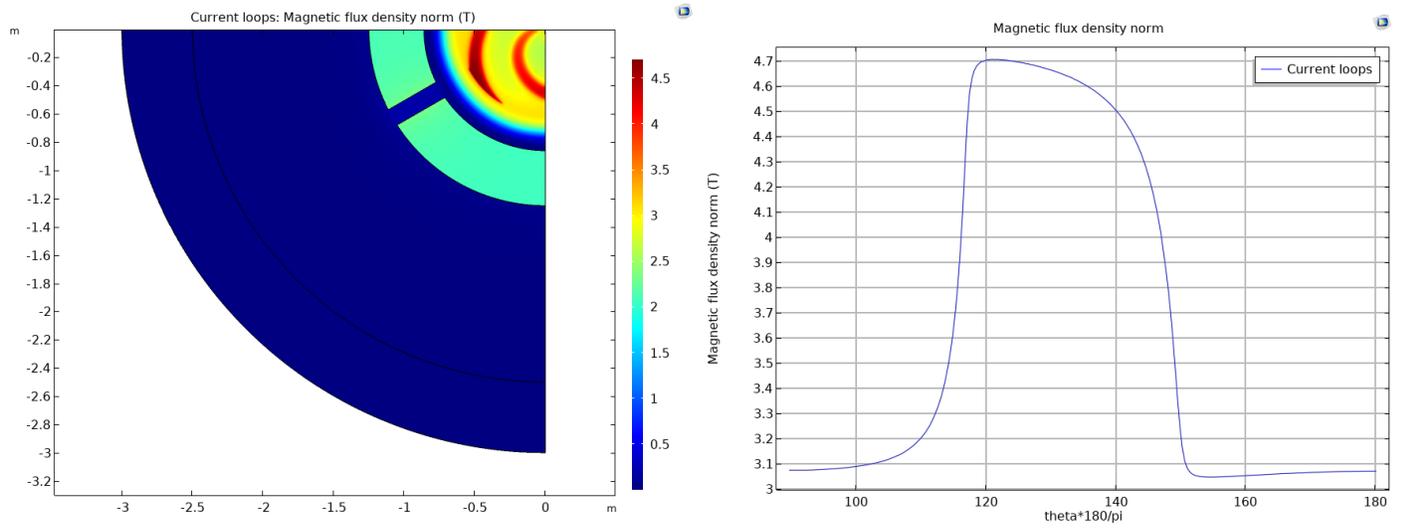

Fig.6: Field distributions for dipole magnet calculated with magnetic vector potential.

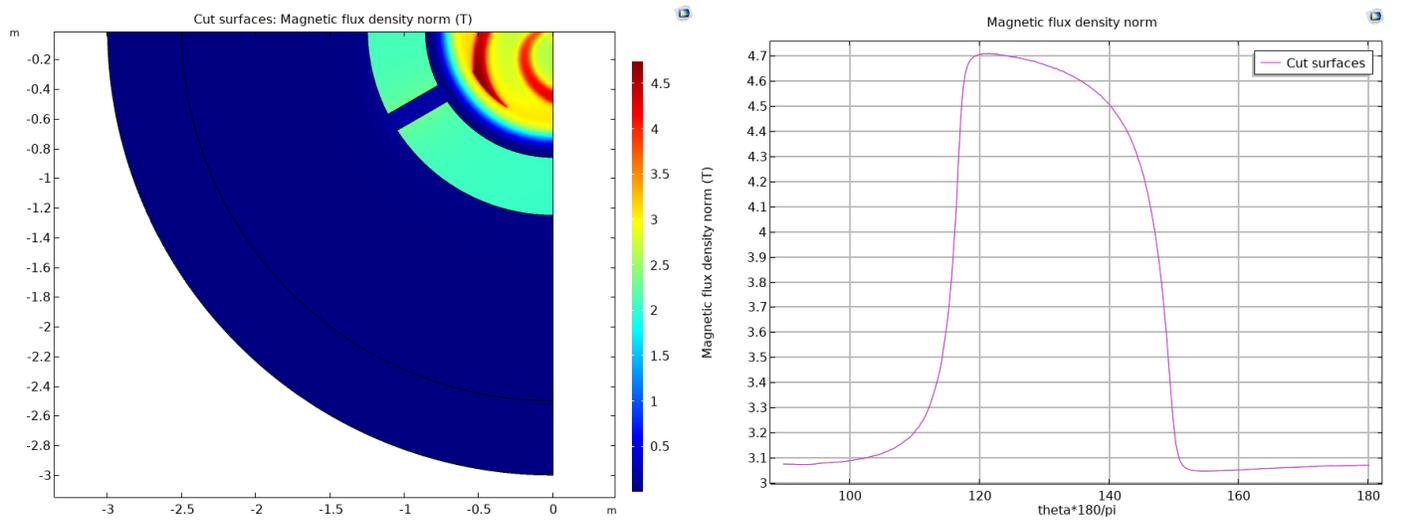

Fig.7: Field distributions for dipole magnet calculated with total scalar potential.

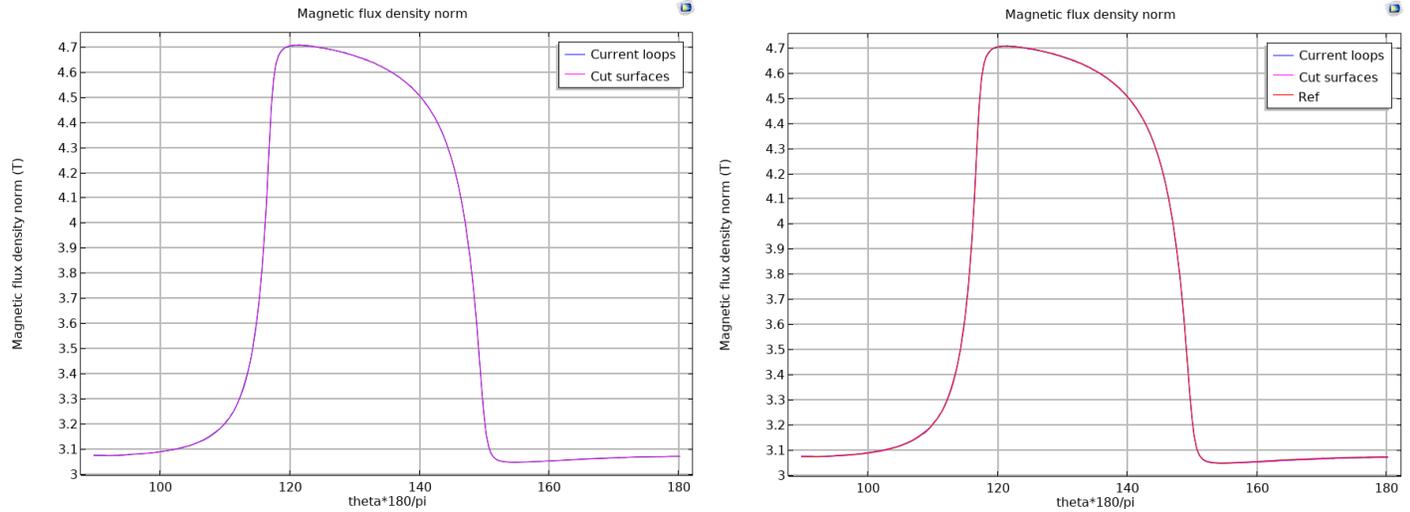

Fig.8: Comparison of the azimuthal profiles with each other and against the reference field.

### 3.2 Calculation of azimuthally averaged radial field distributions

Figures 9 – 11 present the field maps on median plane and the azimuthally averaged profiles along the radius calculated for dipole magnet by the finite-element method using COMSOL Multiphysics®. The reference distributions are shown in Fig.9 and the distributions calculated with magnetic vector and total scalar potentials in Fig.10 and Fig.11, respectively. Figure 12 shows a comparison of the azimuthally averaged radial profiles calculated by the two methods with each other and against the reference field.

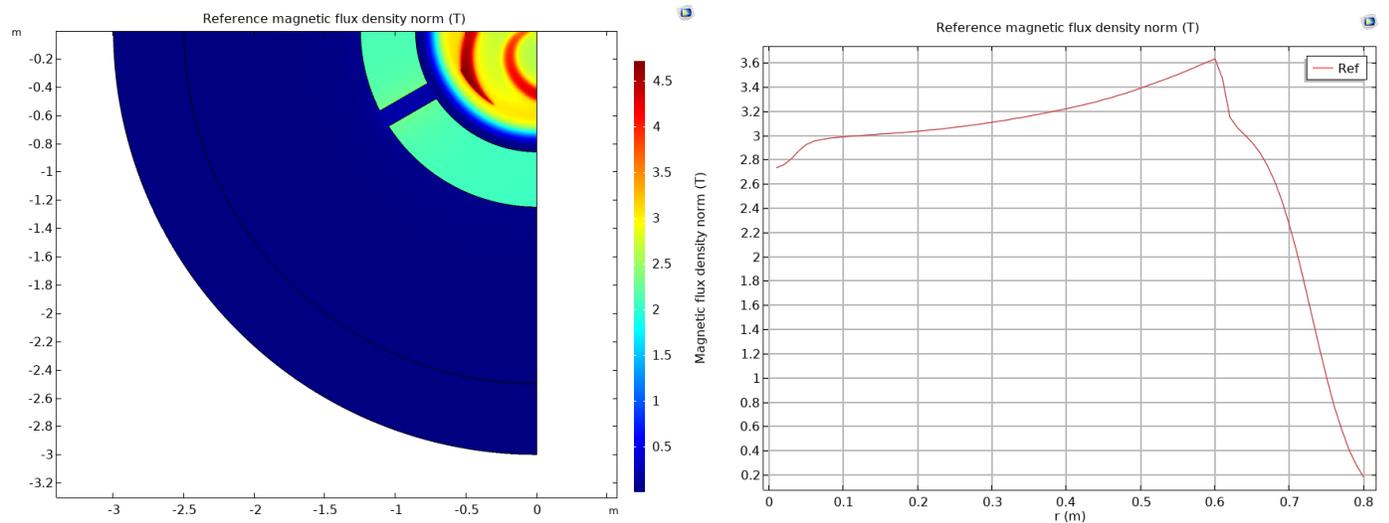

Fig.9: Reference field distributions for dipole magnet.

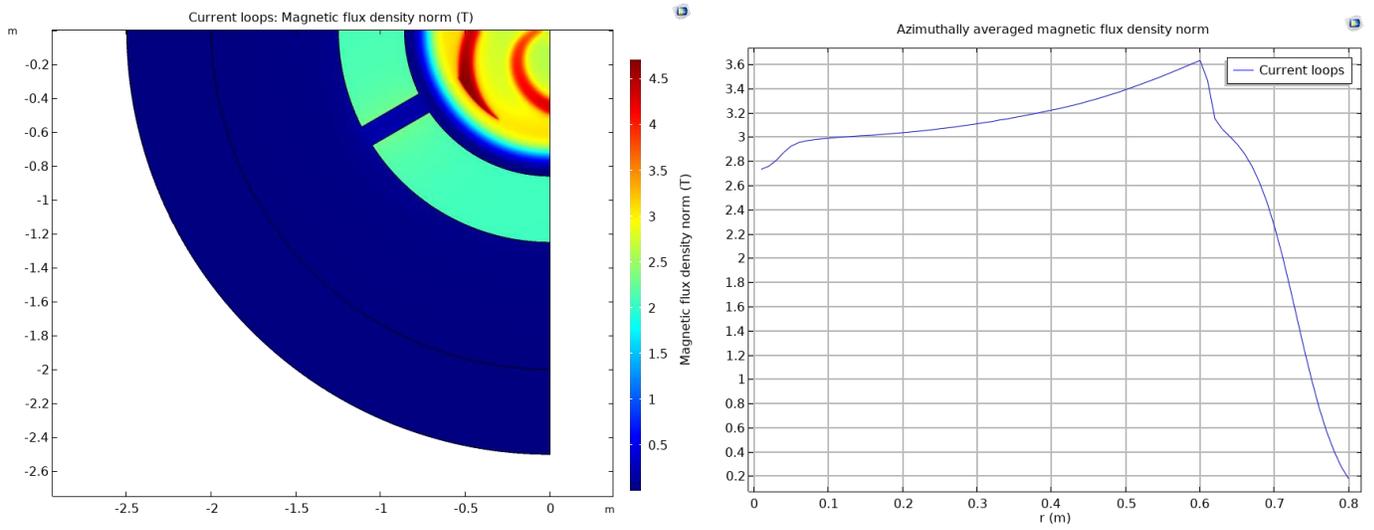

Fig.10: Field distributions for dipole magnet calculated with magnetic vector potential.

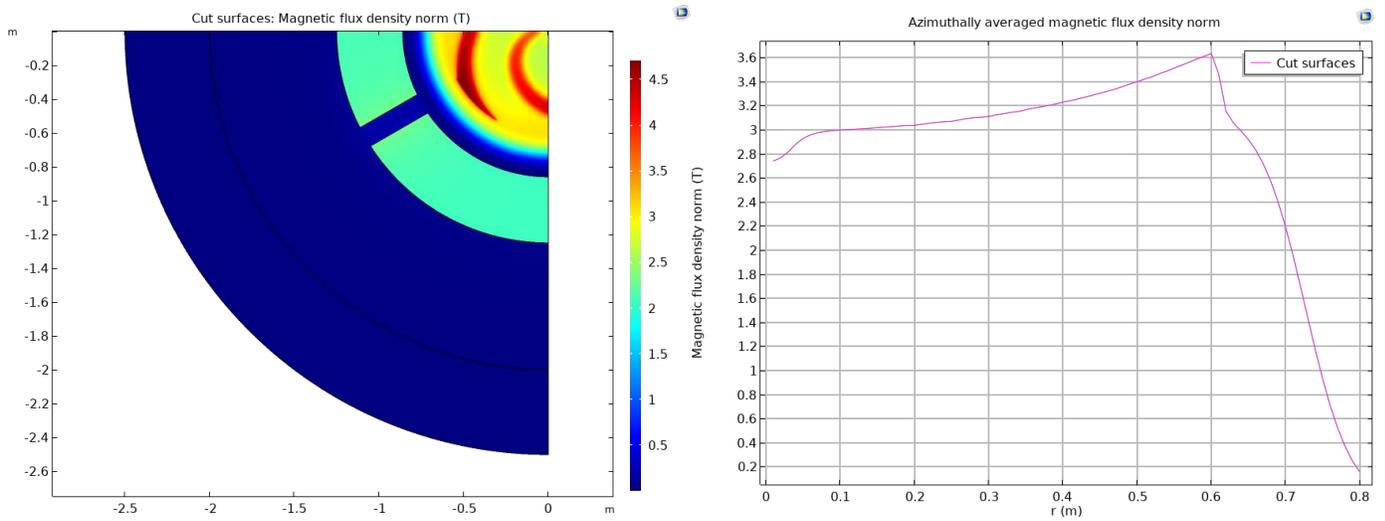

Fig.11: Field distributions for dipole magnet calculated with total scalar potential.

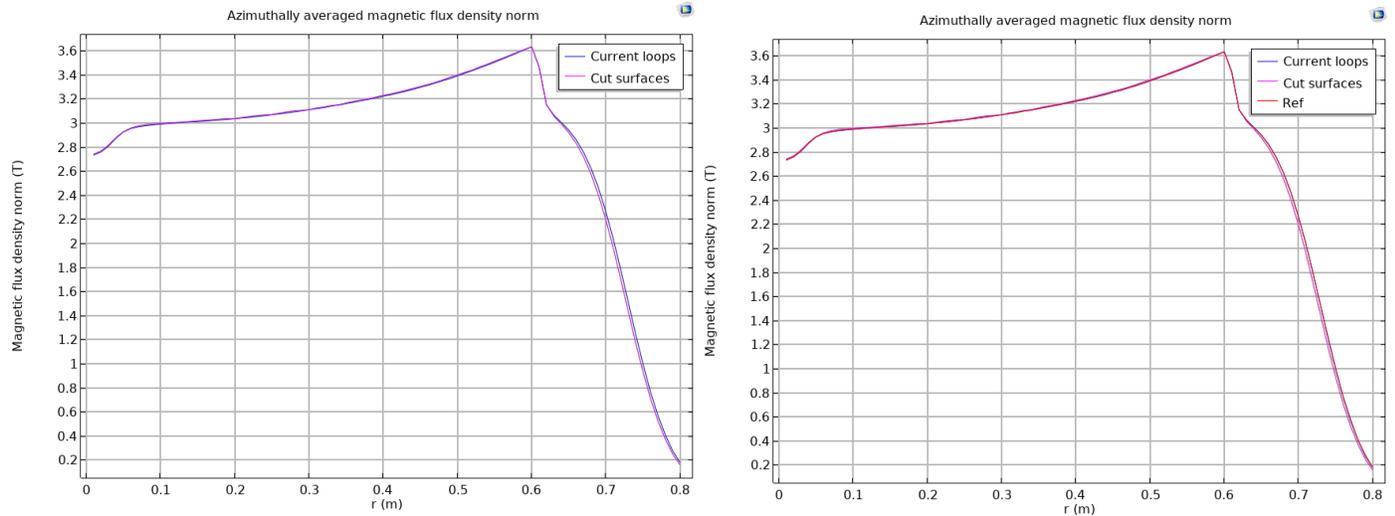

Fig.12: Comparison of the radial profiles with each other and against the reference field.

4. Comparison of the modeling performance

Field simulations in terms of the magnetic vector potential as well as the total scalar potential have been performed on the Intel® Core® i9-9900K CPU @ 3.60 GHz, 64 GB RAM computer by using the COMSOL Multiphysics® v. 5.6 software. As shown in Figures 5 – 12, both computational frameworks produce the expected solutions in excellent qualitative and quantitative agreement. Their performance is analyzed and compared below in terms of the relevant FEM parameters.

For calculation of the azimuthal field distributions, the benchmark geometry of the magnet is approximated by a finite-element mesh consisting of 960 090 elements with minimum quality of 14.7 %. The mesh quality is optimized for the best possible balance between the stability of solution and the computational cost. To improve the calculation precision, both potentials are approximated by third-order (cubic) Lagrange shape functions assigned to the mesh elements. Under such conditions, it requires 18 458 724 degrees of freedom (DOFs) for calculations with magnetic vector potential and 3 774 955 degrees of freedom (DOFs) for calculations with total scalar potential. The usage of computational resources amounts to 49.45 Gb/54.57 Gb of the physical/virtual memory (RAM) and 38 402 s (10h 40m 2s) of the computational time for calculations with magnetic vector potential, but only to 14.76 Gb/17.12 Gb of the physical/virtual memory (RAM) and 3 135 s (52m 15s) of the computational time for calculations with total scalar potential.

The radial field distributions are calculated similar but now for the whole array of points on the radial interval with each partial result being averaged over the azimuth. To mitigate with the expected long-run of calculations without much compromising on the precision, we use second-order (quadratic) Lagrange shape functions for both potentials together with finer mesh for the benchmark geometry of the magnet. The mesh consists of 1 149 877 elements with minimum quality of 16.3 %. Under such conditions, it requires 16 785 021 degrees of freedom (DOFs) for calculations with magnetic vector potential and 5 136 088 degrees of freedom (DOFs) for calculations with total scalar potential. The usage of computational resources amounts to 40.99 Gb/47.50 Gb of the physical/virtual memory (RAM) and 158 579 s (1d 20h 2m 59s) of the computational time for calculations with magnetic vector potential and only to 5.45 Gb/6.28 Gb of the physical/virtual memory (RAM) and 8 660 s (2h 24m 20s) of the computational time for calculations with total scalar potential.

Compared to the reference field, the calculations of the azimuthal field distributions using both potentials produce an excellent agreement with maximum relative error of only 0.1 %. The results are however achieved with significantly different usage of the memory and computation speed. Obviously, the calculation in terms of the total scalar potential is much faster and more economic. Indeed, the number of DOFs per each tetrahedron used in calculations is 45 for third-order edge elements and only 20 for the same order nodal elements.

The calculations of the radial field distributions on the other hand, are less accurate as compared to the reference magnetic field. The maximum relative error increases to 0.45 % for calculations with vector potential and 0.94% for calculations with scalar potential due to the use of the lower order elements and applying an additional averaging procedure over the azimuth. The accuracy of these calculations can be improved by either refining the mesh, or increasing the polynomial order of the shape functions. Both options will require more computational capacity [6].

Conclusion

Finite-element simulations of magnetostatic fields are performed in terms of magnetic vector and total scalar potentials and compared for purpose of modeling the accelerator magnets. The potentials represent the unknown variables associated with the A– and V– formulations of the magnetostatics to describe the magnetic fields. The simulations are carried out with a single software package, the COMSOL Multiphysics®, where both formulations are implemented [2]. The caveats of the two methods for simulations of accelerator magnets are outlined. In particular, the use of vector potential (A–formulation) for the whole problem domain including the extended current-free regions substantially increases the total number of degrees of freedom (DOFs) and the associated computational cost. On the other hand, the use of total scalar potential (V–formulation) requires the supplementary construction of the cut surfaces assigned with the potential jumps for consistency of calculations. The numerical performance of these methods is illustrated with the model example of a superconducting dipole magnet [3] recently developed for operation in the isochronous cyclotron SC200 [4, 5]. To benchmark this model with the two methods, each coil of the magnet is approximated by the line currents located at determined conductor positions inside the coil for computations with magnetic vector potential. The associated potential surfaces bounded by these loops are then constructed and implemented for computations with total scalar potential. All dimensions of the current loops and the potential surfaces are made compatible with each other and preliminary optimized [8]. This provides the uniform geometry of the magnet for both computational methods. The quantity of interest is the magnetic flux in the aperture region in between the poles of the magnet. In this region, the field maps on the median plane as well as the azimuthal and the radial magnetic flux density distributions are derived. The aperture fields are calculated with the precision capable of detecting the harmonic distortion factors of the order of $10^{-4}$ of the field magnitude. The results of calculations are analyzed and compared in terms of the relevant FEM parameters accounting for performance of computation and computational cost. In particular, it is shown that the use of scalar potential as compared to its vector counterpart substantially reduces the number of degrees of freedom, the usage of computer memory, and especially, the computational time, while a similar accuracy of magnetic fields is attained with the both methods. For calculations of the azimuthal field distributions with both potentials the obtained results are in excellent agreement with the reference magnetic field. For calculations of the radial field distributions, the results become less accurate due to the use of lower order finite elements and applying an additional averaging procedure over the azimuth. In conclusion, both potentials are well-suited for finite-element simulations of the nonlinear magnetostatic problems. The total scalar potential can be used to quickly access the results of the modeling on the computers with limited computational capacities. The use of magnetic vector potential on the other hand allows to achieve the desired accuracy of solution when the computational cost is not a decisive factor.